\documentclass[tightenlines,twocolumn,aps,nofootinbib]{revtex4}

\usepackage{psfig,axodraw,float}
\usepackage{graphics}

\def\MSbar{\overline{\mathrm{MS}}}

\def\Li{{\rm Li}}
\def\GEV{{\rm GeV}}

\def\mMS{\overline{m}}
\def\DDiii{\left[{\ln^3(1-z)\over 1-z}\right]_+}
\def\DDii{\left[{\ln^2(1-z)\over 1-z}\right]_+}
\def\DDi{\left[{\ln(1-z)\over 1-z}\right]_+}
\def\DDo{\left[{1\over 1-z}\right]_+}

\begin{document}


\title{

\begin{flushright}
\vbox{
\begin{tabular}{l}
\small UH-511-1070-05\\
\   \end{tabular} }
\end{flushright}

The photon energy spectrum in $B\to X_s + \gamma$ in perturbative QCD through
${\cal O}(\alpha_s^2)$
}

\author{
Kirill Melnikov
$\!\!$\thanks{e-mail:  kirill@phys.hawaii.edu} and
Alexander Mitov
$\!\!$\thanks{ e-mail:  amitov@phys.hawaii.edu}}

\affiliation{Department of Physics and Astronomy,\\
University of Hawaii,\\
Honolulu, HI, 96822}

\begin{abstract}

\vspace{2mm}

We derive the dominant part of the ${\cal O}(\alpha_s^2)$
correction to the photon energy spectrum in the inclusive decay
$B\to X_s+\gamma$. The detailed knowledge of the spectrum is
important for relating the theoretical calculations of the  $B\to
X_s + \gamma$ decay rate and the experimental measurements where
{\rm a cut} on the photon energy is applied. In addition, moments
of the photon energy spectrum are used for the determination of
the $b$-quark mass and other fundamental parameters of heavy quark
physics. Our calculation reduces the theoretical uncertainty
associated with uncalculated higher orders effects and shows that,
for $B\to X_s+\gamma$, QCD radiative corrections to the photon
energy spectrum are under theoretical control.

\end{abstract}

\maketitle

\thispagestyle{empty}

\section{Introduction}

The process $B\to X_s + \gamma$ plays an important role in
particle physics. Within the Standard Model, it is loop-induced;
in many extensions of the Standard Model the decay rate of $B\to
X_s + \gamma $ receives sizable contributions from virtual, yet
undiscovered particles. Comparison of  the experimentally measured
rate for $B\to X_s + \gamma $ with the theoretical expectations
puts stringent constraints on various new physics scenarios.

Thanks to BaBar, Belle and CLEO experiments, high quality data on
$B\to X_s + \gamma$ is available at present
\cite{Belle2004,BaBar,Cleo,HFAG}. Due to the large irreducible
background at small photon energies, all measurements are
performed with a lower cut on the photon energy $E_{\gamma} >
E_{\rm cut}$. Until recently, $E_{\rm cut}\sim 2.0~\GEV$ was
routinely used. Last year Belle \cite{Belle2004} presented precise
measurements of the branching fraction and the first two moments
of the photon energy spectrum with the cut as low as $E_{\rm cut} =
1.815~\GEV$. This year, preliminary results with $E_{\rm cut} \ge
1.9~\GEV$ were reported by BaBar collaboration \cite{BaBar}.

The theoretical understanding of the decay  $B\to X_s+\gamma$ in
the Standard Model is currently quite robust. The total decay rate
is known through next-to-leading order (NLO) in perturbative QCD
\cite{rate1,rate2,n1,rate3,n2,n3,n4,rate4,rate5,n5,rate6,rate7}
and the process of calculating it through next-to-next-to-leading
order (NNLO) is under way
\cite{totrate1,totrate2,totrate3,totrate4,totrate5}. However,
because the experimental measurements require a cut on $E_\gamma$,
even a perfect knowledge of the total, fully inclusive decay rate
is only useful if the the photon energy spectrum is understood
well enough to relate a  measurement of the branching fraction
with $E_{\gamma} > E_{\rm cut}$ to the theoretically known
inclusive total rate. In addition, the detailed knowledge of the
photon energy spectrum has its own merit. Since the photon energy
spectrum is largely insensitive to contributions from beyond the
Standard Model physics, it is used to study strong interaction
physics in heavy flavor decays. In particular, the value of the
$b$-quark mass, the average kinetic energy of a $b$ quark in a
$B$-meson and the importance of the shape function for different
values of the cut on the photon energy can be investigated.

The theoretical description of the radiative $b$-decays is based
on an effective theory approach where all particles heavier than
the $b$ quark are integrated out. The full effective Lagrangian
that is used in the calculation of radiative decays of $B$-mesons
can be found in \cite{Heff1,rate4}. It is well established
\cite{KN,LLMW} that the dominant contribution to the photon energy
spectrum comes from the local operator $\hat{O}_7 \sim \bar s
\sigma_{\mu \nu} b F_{\mu \nu} $, while for the computation of the
total decay rate, other operators are also important.

The photon energy spectrum in $b \to X_s + \gamma$ is currently
known through ${\cal O}(\alpha_s)$. The effects of higher order
QCD corrections are traditionally studied \cite{LLMW} in the
so-called Brodsky-Lepage-Mackenzie (BLM) approximation \cite{BLM};
the BLM corrections are both the easiest to compute and often
provide the dominant part of the ${\cal O}(\alpha_s^2)$ corrections. In flavor
physics, where  typical energy scales are relatively low and
hence the strong coupling constant is large, the BLM corrections
can be significant and their naive application could lead to
inflated theoretical uncertainties. A consistent application of the
BLM corrections in $B$-decays was developed in a series of papers
\cite{kolya,kolya1}. Similar approach to the photon energy
spectrum in $B \to X_{s} + \gamma$ has been initiated in
\cite{shifman} and later elaborated upon in \cite{BU2002,BBU2004}
where, in particular, the BLM corrections have been resummed to
all orders in the strong coupling constant. Among other things,
this approach requires a careful separation of perturbative and
non-perturbative effects and understanding  the intricate
interplay between them.

Additional theoretical information on the shape of the photon
energy spectrum or on the integrated spectrum in the  presence of a
lower cut on the photon energy, can be obtained from the
universality of the soft and collinear gluon radiation
\cite{ratin}. Such an approach can be used to predict large
logarithms $\sim\ln(E_{\rm max}-E_{\rm cut})$, where $E_{\rm max}$
is the maximal energy allowed for the photon in $B \to X_s +
\gamma$. At present, such computations are performed with
next-to-next-to-leading logarithmic accuracy \cite{GardiAnd} (see
also \cite{AR}). On general grounds it is clear that such an
approach is applicable only for $E_{\rm cut}/E_{\rm max}\approx 1$
and not for moderate or low values of the cut on the photon
energy.

Further improvement in the theoretical description of the photon
energy spectrum in $b \to X_s + \gamma$  requires ${\cal
O}(\alpha_s^2)$ corrections beyond the BLM approximation and it is
the purpose of this paper to provide them. We restrict ourselves
to the contribution generated by the operator $\hat O_7$ since, as
we pointed out above, this operator provides the leading
contribution to the photon energy spectrum.

The paper is organized as follows. In Section II we present the
${\cal O}(\alpha_s^2)$ correction to the photon energy spectrum in
$b\to X_s+\gamma$ and discuss its derivation. In Section III we
study the properties of the spectrum and compare it with known
partial results. In Section IV we consider the effect of our
result on the evaluation of the first moment of the photon energy
spectrum and on the branching ratio ${\rm Br}[B\to X_s \gamma]$ in
the presence of a lower cut on the photon energy. Finally, we
present our conclusions.

\section{The perturbative evaluation of the photon spectrum}

We consider the decay of an on-shell $b$-quark into a photon and
any hadronic state containing a strange quark. We assume that this
decay is mediated by the effective operator $\hat{O}_7$:
\begin{equation}
\hat{O}_7={e\over 32\pi^2}\mMS~\overline{s}\sigma_{\mu\nu}
F^{\mu\nu}(1+\gamma_5) b.  \label{O7}
\end{equation}
Here, $F_{\mu\nu}$ is the electromagnetic field strength tensor,
$e = \sqrt{4\pi \alpha}$, where $\alpha$ is the fine structure
constant evaluated at zero momentum transfer \cite{CM} and $\mMS$
is the $\MSbar$ $b$-quark mass evaluated at the $b$-quark pole
mass $\mMS=\mMS(m_b)$. The photon energy spectrum is parameterized
in terms of the variable:
\begin{equation}
z  = {2p_\gamma p_b\over m_b^2} = \frac{2E_\gamma}{m_b}, \label{z}
\end{equation}
where the last equality is valid in the $b$-quark rest frame.
When the photon energy spectrum is studied within perturbation
theory, the kinematically allowed range for the photon energy
implies\footnote{Because of the Fermi motion of the heavy
quark inside the $B$-meson, the photon energy spectrum extends
beyond the point $m_b/2$ all the way up to $E_{\rm max} =
M_{B}/2$, where $M_B$ is the mass of the $B$-meson. }
$0\leq z\leq 1$. For our calculation, we neglect all quark masses
other than the mass of the $b$ quark and use dimensional
regularization for ultraviolet, infra-red and collinear
singularities. Because we restrict ourselves to the operator $\hat
O_7$, the photon energy spectrum is finite after the standard
renormalization procedure is performed.

At leading order in perturbation theory the kinematics of the
decay $b \to s + \gamma$ is very simple: the photon and the
(massless) strange quark are back-to-back with equal energy and
momentum. Such a simple kinematics implies that the energy
spectrum of the photon at this order is given by a delta-function
$\sim\delta(1-z)$. At order ${\cal O}(\alpha_s)$, the
gluon radiation smears the photon energy spectrum and a radiative
tail down to $E_\gamma = 0$ appears. Therefore, the
${\cal O}(\alpha_s^2)$ correction to
the photon energy spectrum is the first
non-trivial QCD correction to the radiative tail.

For the physical observables studied in this paper, we
only need to  compute the ${\cal O}(\alpha_s^2)$
corrections to the photon energy spectrum away from the
maximal value of the photon energy $E_\gamma < E_{\rm max}$. As a
consequence, we do not compute the two-loop virtual
corrections to $b \to s + \gamma$ since they only contribute for
$z = 1$. Therefore, to calculate the ${\cal O}(\alpha_s^2)$
corrections to the photon energy spectrum in $b \to X_s + \gamma$
for $z<1$, we have to consider processes with up to two gluons or
a quark-antiquark pair in addition to the strange quark and the
photon in the final state, as well as the virtual corrections to
single gluon emission. To perform the calculation, we use the
techniques introduced in \cite{AM,ADM,ADMP}. The idea is to apply
the optical theorem to $b \to b$ transition amplitude in the
presence of the constraint on the photon energy. The
constraint, $\delta(E_\gamma - z E_{\rm max})$, is treated as the
on-shell condition for a ``fake'' particle. Then,  evaluation
of the complicated integrals over multi-particle phase-space, e.g.
$b \to s + \gamma + g + g$ is simplified by multiloop integration
technology \cite{tkachov}. To solve the integration-by-parts identities, we
use the algorithm of Ref.~\cite{Laporta} implemented in
\cite{babis}.

It is convenient to present the result for the photon energy
spectrum in a normalized form, dividing the differential rate
${\rm d}\Gamma/{\rm d}z$ by the total width for $B \to X_s +
\gamma$. This ratio is convenient because its integral over the
fraction of the photon energy $z$ is equal to one to all orders in the
strong coupling constant $\alpha_s$:
\begin{equation}
\int \limits_{0}^{1} \frac{1}{\Gamma} {d\Gamma \over {dz} } dz = 1 .
\label{normalization}
\end{equation}
Note that because of Eq.(\ref{normalization}), we can restore the
$\delta(1-z)$ terms in the normalized photon energy spectrum
without explicit computation of the two-loop virtual corrections.

Our result for the photon energy spectrum reads:
\begin{equation}
\frac{1}{\Gamma} {d\Gamma \over  dz} =
\delta(1-z) +  {\left ( \alpha_s\over \pi \right )}~ C_F F^{(1)}
+ \left({\alpha_s\over \pi}\right)^2 C_F F^{(2)},
\label{result}
\end{equation}
where
\begin{equation}
F_2 =  C_FF^{(2,{\rm a})}
+C_A F^{(2,{\rm na})} + T_Rn_fF^{(2,{\rm cf})},
\nonumber
\end{equation}
$\alpha_s$ is the ${\overline {\rm MS}}$ coupling constant
renormalized at the $b$-quark pole mass, $C_F = 4/3,~C_A = 3, T_R
= 1/2$. Also, $n_f = 4$ is the number of light fermion flavors.

The coefficient functions read:
\begin{eqnarray}
&& F^{(1)} = \Bigg\{-{31\over 12}\delta(1-z) -\DDi
-{7\over 4}\DDo
\nonumber \\
&& -{z+1\over 2}\ln(1- z)
+{7+z-2z^2 \over 4}\Bigg\},
\end{eqnarray}
and
\begin{widetext}
\begin{eqnarray}
&& F^{(2,{\rm a})} =
S_{\rm a}\delta(1-z)+
\frac{1}{2}\DDiii+\frac{21}{8}\DDii+\left(-\frac{\pi^2}{6}+\frac{271}
{48}\right)\DDi
\nonumber\\
&&+\left(\frac{425}{96}-\frac{\pi^2}{6}-\frac{\zeta(3)}{2}
\right)\DDo+\frac{4z-4z^2+1+z^3}{2(z-1)} \left[
\Li_3\left(\frac{z}{2-z}\right) - \Li_3\left(-\frac{z}{2-z}\right) -
2\Li_3\left(\frac{1}{2-z}\right) +\frac{\zeta(3)}{4} \right]
\nonumber\\
&&+\left[\frac{z^3-2z^2+2z-3}{2(z-1)}\ln(1-z)-
\frac{-140z^4+219z^3-124z^2+28z+27z^5+9z^6+z^8-6z^7-6}{12z(z-1)^3}\right]
\Li_2(z-1)\nonumber\\
&&-2(z-1)^2\Li_3(z-1)+
\left[ \frac{2z^3-9z^2-2z+11}{4(z-1)}\ln(1-z)-\frac{-27z^2+8z^6-9+21z-3z^3+64z^4-
46z^5}{12z(z-1)^3}\right]\Li_2(1-z)\nonumber\\
&&-\frac{-17z^2+4z+4z^3+11}{4(z-1)}\Li_3(1
-z)-\frac{2z^3+13-9z^2}{4(z-1)}\Li_3(z)+\frac{4z-4z^2+1+z^3}{6(z-1)}\ln^3(2-z
)\nonumber\\
&&+\left[-\frac{4z-4z^2+1+z^3}{2(z-1)}\ln^2(1-z)-\frac{-140z^4+219z^3-124z^2+
28z+27z^5+9z^6+z^8-6z^7-6}{12z(z-1)^3}\ln(1-z)\right.\nonumber\\
&&\left. -\frac{4z-4z^2+1+z^3}{z-1}\frac{\pi^2}{12}\right]\ln(2-z)
+\frac{z^3-2z^2+2z+1}{4z}\ln^3(1-z)+
\frac{z^5-3z^4+5z^3+7z^2+5z-9}{24z}\ln^2(1-z)\nonumber\\
&&+\left[-\frac{z^2+8z-11}{8(z-1)}\ln^2(1-z)-\frac{-27z^2+8z^6-9+21z-3z^3
+64z^4-46z^5}{12z(z-1)^3}\ln(1-z)\right]\ln(z)\nonumber\\
&&+\left[(-z^2+z-3)\frac{\pi^2}{12} -
\frac{4z^5+151z+2z^4-48z^2-41z^3-36}{48z(z-1)}\right]\ln(1-z)-
\frac{(z-2)(z^4-z^3-11z^2+13z+3)}{z}\frac{\pi^2}{72}\nonumber\\
&&+\frac{z^3-11z^2-2z+18}{4(z-1)}\zeta(3)-\frac{8z^4-244z^3+
175z^2+598z-569}{96(z-1)} ~ , \label{CFCF}
\end{eqnarray}
\end{widetext}
\vskip 1mm
\begin{widetext}
\begin{eqnarray}
&& F^{(2,{\rm na})} =
S_{\rm na}\delta(1-z)+\frac{11}{8}\DDii+\left(\frac{\pi^2}{12}+
\frac{95}{144}\right)\DDi+\left(
\frac{\zeta(3)}{4}-\frac{905}{288}+\frac{17\pi^2}{72}\right)\DDo
\nonumber\\
&&-\frac{4z-4z^2+1+z^3}{4(z-1)}\left[
\Li_3\left(\frac{z}{2-z}\right) -\Li_3\left(- \frac{z}{2-z}\right)
-2\Li_3\left(\frac{1}{2-z}\right) +\frac{\zeta(3)}{4}\right] +(z-1
)^2\Li_3(z-1) \nonumber\\
&&+\left[-\frac{z^3-2z^2+2z-3}{4(z-1)}\ln(1-z)+\frac{-140z^4+219z^
3-124z^2+28z+27z^5+9z^6+z^8-6z^7-6}{24z(z-1)^3}\right]\Li_2(z-1)
\nonumber\\
&&+\left[\frac{z(3-z)}{4}\ln(1-z)+\frac{(1+z)
(2z^4-29z^3+73z^2-57z+15)}{24(z-1)^3}\right]\Li_2(1-z)
+\frac{4z-4z^2+1+z^3}{4(z-1)}\Li_3(z)
\nonumber\\
&&+\frac{(z-3)z}{2}\Li_3(1-z)-\frac{4z-4z^2+1+z^3}{12(z-1)}\ln^3(2-z)
+\left[\frac{4z-4z^2+1+z^3}{4(z-1)}\ln^2(1-z)
\right.\nonumber\\
&& \left.
+\frac{-140z^4+219z^3-124z^2+28z+27z^5+9z^6+z^8-6z^7-6}{24z(z-1)^3}
\ln(1-z) +\frac{4z-4z^2+1+z^3}{z-1}\frac{\pi^2}{24}\right]\ln(2-z)
\nonumber\\
&&+\frac{(1+z)(2z^4-29z^3+73z^2-57z+15)}{24(z-1)^3}\ln(1-z)\ln(z)
-\frac{(z-1)^2}{8}\ln^3(1-z)-\frac{(z+2)(z^3-5z^2+9z-35)}{48}\ln^2(1-z)
\nonumber\\
&& +\left[(z^2-z+3)\frac{\pi^2}{24}
+\frac{6z^5+72-392z^3+51z^4+219z^2+92z}{144z( z-1)}
\right]\ln(1-z)+\frac{z^5-3z^4-3z^3+34z^2-24z+3}{z}\frac{\pi^2}{144}
\nonumber\\
&&-\frac{z^3-10z^2+6z+7}{8(z-1)}\zeta(3)+
\frac{12z^4-754z^3+1191z^2+264z-761}{288(z-1)} ~ ,\label{CACF}
\end{eqnarray}
\end{widetext}

\vskip 2mm

\begin{widetext}
\begin{eqnarray}
&& F^{(2,{\rm nf})} = S_{\rm nf}\delta(1-z)
-\frac{1}{2}\DDii-\frac{13}{36}\DDi+\left(-\frac{\pi^2}{18}
+\frac{85}{72}\right)\DDo
\nonumber\\
&&+\frac{z^2-3}{6(z-1)}\Li_2(1-z)+
\frac{z^2-3}{6(z-1)}\ln(1-z)\ln(z)-\frac{1+z}{4}\ln^2(1-z)-
\frac{6z^3-25z^2-z-18}{36z}\ln(1-z) \nonumber\\
&& -(1+z)\frac{\pi^2}{36} +\frac{-49+38z^2-55z}{72} ~ .
\label{CFTRnf}
\end{eqnarray}
\end{widetext}
Here, ${\rm Li}_3(x) = \int \limits_{0}^{x} {\rm d}x_1 {\rm
Li}_2(x_1)/x_1$, $\zeta(3)$ is the Riemann zeta-function and
$[\ln^n(1-x)/(1-x)]_+$ are the plus-distributions defined in the
standard way. The constants $S_{i}$ take the following values $
S_{{\rm a}}= 1.216 ,~S_{\rm na }= -4.795 ,~ S_{\rm nf}=
49/24+\pi^2/8-2\zeta(3)/3 \approx 2.474$. In addition, for
simplicity, we have set the renormalization and the factorization
scales in Eq.(\ref{result}) equal to the $b$-quark pole mass
$m_b$. The total width of the radiative transition $B \to X_s +
\gamma$ is known
%
%
through ${\cal O}(\alpha_s)$. It reads:
\begin{equation}
\Gamma = \Gamma^{(0)} \left(
1+{\alpha_s\over \pi}C_F\left( {4\over 3}-{\pi^2\over 3}\right) +
{\cal{O}}(\alpha_s^2) \right) ,
\label{Gammatot}
\end{equation}
where $\displaystyle \Gamma^{(0)} = {\alpha G_F^2\over 32\pi^4}
\vert V_{tb}V_{ts}^*\vert^2 C_7^2(m_b) ~m_b^3\mMS^2(m_b)$.

Our result for the photon energy spectrum Eq.(\ref{result}) can be
compared to partial results that exist in the
literature. The correction to the spectrum proportional to the
number of massless flavors $n_f$ agrees with the result in \cite{LLMW}.
In addition, the terms in the photon  energy spectrum that develop
$[\ln^n(1-z)/(1-z)]_+$, $n=0\dots 3$ singularities for $z \to 1$
agree with \cite{GardiAnd}. We note that the photon energy spectrum
computed in this paper provides an additional check
 of the constant $D_2$ that was
extracted in \cite{Gardi} from our previous calculation of the
perturbative heavy quark fragmentation function \cite{Dini}.

\section{Analysis of the Perturbative Spectrum}

In this Section we discuss the properties of the perturbative
photon energy spectrum Eq.(\ref{result}) and compare it with
various approximations available in the literature. For the
illustration we use $\alpha_s = \alpha_s(m_b) = 0.22$ for all
plots in this Section. A choice of $\alpha_s$ that is more
appropriate for observables when a cut on the photon energy is
applied, is discussed in the next Section.

\begin{figure}
\centerline{\resizebox{0.40\textwidth}{!}{\includegraphics{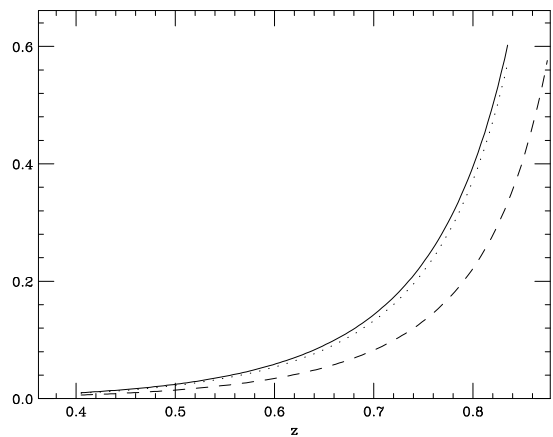}}}
\caption{\small The normalized perturbative spectrum
Eq.(\ref{result}) in $b\to s+\gamma$ retaining  ${\cal
O}(\alpha_s^2)$ (solid), BLM (dots) and ${\cal O}(\alpha_s)$
terms. } \label{q1}
\end{figure}

An overview of the photon energy spectrum is given in
Fig.\ref{q1}, where the ${\cal O}(\alpha_s)$, the full ${\cal
O}(\alpha_s^2)$ and the BLM approximation to the spectrum  are
compared. It follows that when the renormalization scale is
chosen equal to the mass of the $b$-quark,
the bulk of the ${\cal O}(\alpha_s^2)$
correction is provided by the BLM terms and that the
non-BLM corrections are moderate.

As we mentioned in the Introduction, the BLM corrections require a
special treatment; given their magnitude, the naive interpretation
of Fig.\ref{q1} results in a large perturbative uncertainty on the
photon energy spectrum. Fortunately, the effects of the BLM
corrections have been resummed to all orders in a way consistent
with the Wilsonian approach to the operator product expansion
(OPE) for $b \to X_s + \gamma$ \cite{BBU2004}. Since the impact of
the BLM corrections  on the photon energy spectrum is
well-understood, we can concentrate in the following analysis on
the effect of the non-BLM corrections.
This is the new contribution to the photon energy spectrum derived
in this paper.

\begin{figure}
\centerline{\resizebox{0.40\textwidth}{!}{\includegraphics{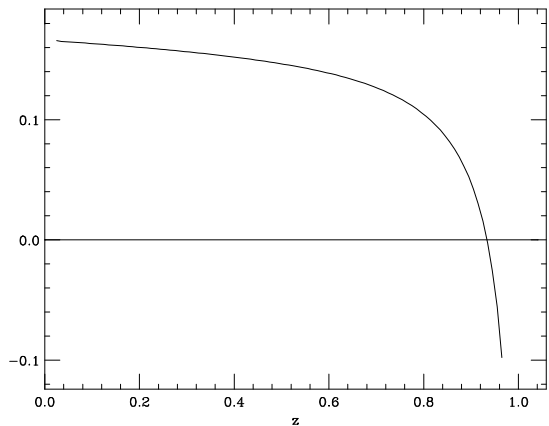}}}
\caption{\small The non-BLM ${\cal O}(\alpha_s^2)$ correction to
the photon energy spectrum relative to the ${\cal O}(\alpha_s)$
correction. } \label{q2}
\end{figure}

The magnitude of the non-BLM part of the ${\cal O}(\alpha_s^2)$
correction is shown in Fig.\ref{q2}, where it is plotted relative
to the photon energy spectrum through  ${\cal O}(\alpha_s)$. The
relative size of the non-BLM ${\cal O}(\alpha_s^2)$ correction is
within $\pm 10 \%$, approximately.
%


%
%

\begin{figure}
\centerline{\resizebox{0.40\textwidth}{!}{\includegraphics{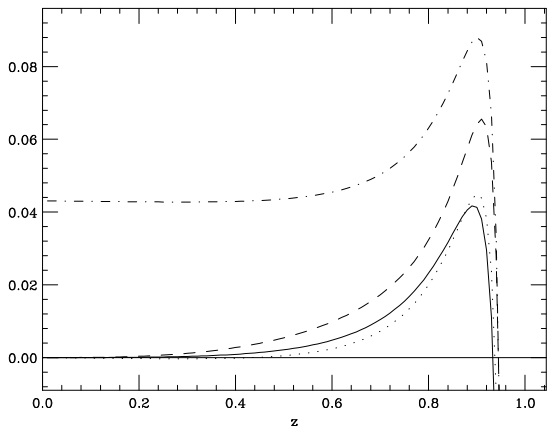}}}
\caption{\small The non-BLM ${\cal O}(\alpha_s^2)$
correction (solid), its $z \to 1$ approximation (dots-dashes), the
$z \to 1$ approximation minus its value for $z=0$ (dots) and
the $z \to 1$ asymptotics times $z^3$ (dashes). }
\label{q4}
\end{figure}

\begin{figure}
\centerline{\resizebox{0.40\textwidth}{!}
{\includegraphics{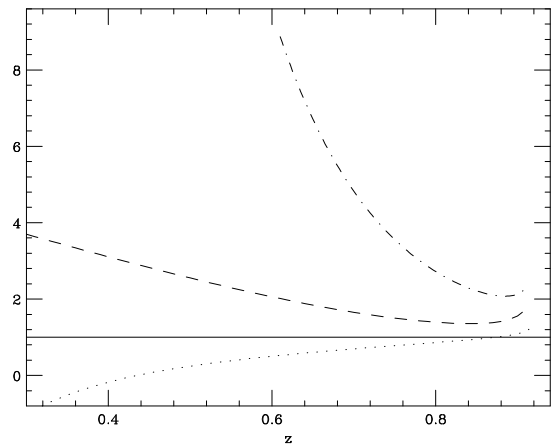}}}
\caption{\small Same as Fig.\ref{q4}. Each curve is normalized
to the non-BLM ${\cal O}(\alpha_s^2)$ correction. }
\label{q5}
\end{figure}

It is interesting to check if the terms
$[\ln^n(1-z)/(1-z)]_+$, $n=0\dots 3$ in Eq.(\ref{result}) that
are singular in $z \to 1$ limit,
provide a good approximation to non-BLM corrections;
Figs.\ref{q4},\ref{q5} illustrate this. We see that
the singular terms  do not furnish a good approximation
to the exact result for moderately large values of $z$; for example,
for $z = 0.8$, the $z \to 1$ approximation overestimates the exact
result by a factor of three. This  comparison suggests that
based on the previously known large $z$ behavior of the
non-BLM piece \cite{GardiAnd}, one could not have computed its
contribution to observables, unless very high value of the cut on the
photon energy is in place.

Although the $z \to 1$ approximation alone is insufficient to
adequately approximate the non-BLM correction, it
is possible to improve the quality of the approximation by
combining the $z \to 1$ approximation with the
$z\to 0$ behavior of
the photon energy spectrum. It is easy to see that
for $\hat{O}_7$ operator
the photon energy spectrum should vanish as $z^3$, for $z \to 0$.
Hence, we can modify the $z \to 1$ asymptotics by multiplying it
by $z^3$. This improves the quality of the approximation; for
$z=0.8$, the modified $z \to 1$ approximation overestimates the
exact result by a factor $1.3$. However, the best description of
the exact result is obtained by subtracting from $z \to 1$
asymptotics its value at $z=0$. While this approximation is hard to
justify theoretically, it gives quite an accurate description of
the spectrum for moderately large values of $z$.

\section{Applications}

In this Section, we study the numerical implications of the ${\cal
O}(\alpha_s^2)$ correction to the photon energy spectrum derived
in this paper. As noted in the Introduction, high-quality
experimental results on $B \to X_s + \gamma$ become available from
the $B$-factories. In particular the branching fraction of $B
\to X_s + \gamma$ and the average
energy of the photon in the presence of a cut
$E_\gamma > E_{\rm cut}$ are measured precisely.

The applicability of the pure perturbative description of the
photon energy spectrum depends upon the exact value of $E_{\rm
cut}$ which, in turn, defines the degree of ``inclusiveness'' of
the process. If $E_{\rm cut}$ is high, the usual OPE breaks down
and the resummation of leading twist effects in the OPE leads to
the appearance of the non-perturbative shape function. By
decreasing $E_{\rm cut}$ it becomes possible to describe the
non-perturbative component of the photon energy spectrum by means
of the local OPE. It is generally believed that for $E_{\rm cut}$
as low as $1.8~{\rm GeV}$, we are already in the OPE regime.
Fortunately, Belle \cite{Belle2004} published results on the
branching fraction and the first two moments for precisely this
value of the cut on the photon energy $E_\gamma > 1.815~{\rm
GeV}$.

To proceed further, we have to
specify the numerical value of the $b$-quark mass.
Although our result for the spectrum is derived in the pole
scheme, it is clear on general grounds that the pole mass can not
appear in short-distance quantities; instead, a consistent
application of OPE leads to short-distance, low-scale quark masses
that enter the perturbative calculations. In principle, the
transformation from one scheme to another should be done in a
self-consistent way, including the proper treatment of
non-perturbative effects. However, as we will see, the
contribution of the non-BLM ${\cal O}(\alpha_s^2)$ correction to
observables is relatively small and, therefore, a simple
estimate suffices. For this reason, instead of the pole
$b$-quark mass we use in our formulas $m_b = 4.6~{\rm GeV}$ which
roughly approximates the kinetic $b$-quark mass normalized at $\mu
= 1~{\rm GeV}$ \cite{prl}. For the two applications that we
consider below, the exact value of the $b$ quark mass mostly
affects the $z$ integration region, through $z > z_{\rm
cut} = 2 E_{\rm cut}/m_b$. In addition, we note that the scale at
which the strong coupling constant has to be evaluated in the
numerical estimates has to be smaller than the mass of the
$b$-quark. This follows from the fact that, for a given value of
$E_{\rm cut}$, the maximal value of the invariant mass of the
hadronic system in $b \to X_s+\gamma$ is $\sqrt{m_b (m_b - 2E_{\rm
cut})} \sim 2~{\rm GeV}$ for $E_{\rm cut} \sim 2~{\rm GeV}$. This
implies that for typical values of the cut on the photon energy
the appropriate value for the scale at which $\alpha_s$ should be
evaluated is $\mu \sim 1.5 - 2~{\rm GeV}$. Having made these
preliminary remarks, we are ready to present our numerical
estimates.

\begin{figure}
\centerline{\resizebox{0.40\textwidth}{!}{\includegraphics{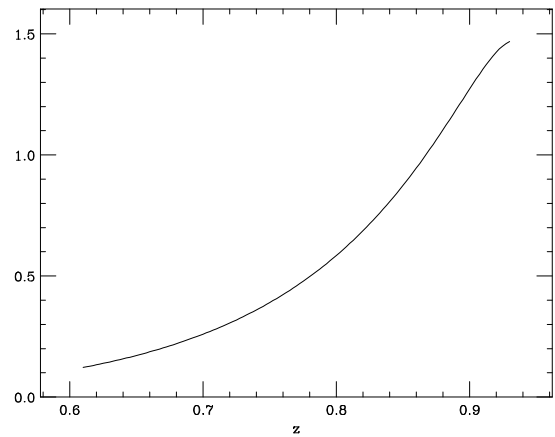}}}
\caption{\small The function $\Delta_R(z)$ Eq.(\ref{R})
for an experimentally relevant range of $z_{cut}$. Only the non-BLM
corrections are used in the calculation. } \label{q6}
\end{figure}

First, we consider the fraction of events that contain a photon
with energy above the cut $E_{\rm cut}$. The second order QCD
corrections shift this fraction by:
\begin{eqnarray}
&&  \delta R(E_{\rm cut}) =  - \left ( \frac{\alpha_s}{\pi} \right
)^2 \Delta_R(z_{\rm cut}), \nonumber \\
&& \Delta_R(z_{\rm cut}) = C_F \int \limits_0^{z_{\rm cut}} {\rm
d}z F^{(2)}(z) , \label{R}
\end{eqnarray}
where we have used Eq.(\ref{normalization}) to express the
integral over $z > z_{\rm cut}$ through the lower part of the
photon energy spectrum. As we explained previously, the
BLM corrections are taken into account in the estimates for
the fraction of events that exist in the  literature.
Therefore, we disregard the BLM corrections and
present the contribution of the non-BLM
corrections to Eq.(\ref{R}) in Fig.\ref{q6}.
With our choice of the $b$-quark mass and for $E_{\rm cut} = 1.8~{\rm GeV}$,
we find $z_{\rm cut} = 0.78$ and $\Delta_R= 0.50$ which, depending
on the value of $\alpha_s = 0.3
\frac{.}{\cdot} 0.35$, translates into:
\begin{eqnarray}
\delta R_{\rm non-BLM} = -( 5.5 \pm 0.5 ) \times 10^{-3}.\nonumber
\end{eqnarray}

The predictions for the fraction of events with
the photon energy larger than $1.8~{\rm GeV}$ that are available
in the literature are:
$R = 0.952^{+0.013}_{-0.029}$ \cite{KN}, $R =
0.958^{+0.013}_{-0.029}$ \cite{rate5}, $R = 0.95 \pm 0.01$
\cite{BU2002} and $R = 0.89^{+0.06}_{-0.07}$ \cite{Neubert2004},
where in the last result only perturbative uncertainty is displayed.
We see that the second order non-BLM QCD
corrections change the fraction of events with $E_\gamma > E_{\rm
cut}$ by less than a percent and are about half of the uncertainty
assigned to $R(E_{\rm cut})$ in \cite{BU2002}.

\begin{figure}
\centerline{\resizebox{0.40\textwidth}{!}{\includegraphics{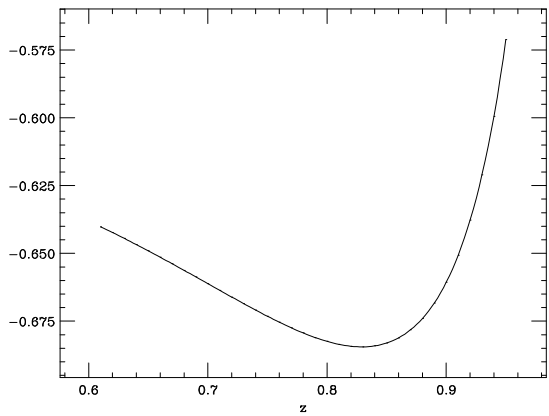}}}
\caption{\small The function $\Delta_E(z)$ Eq.(\ref{E1})
for an experimentally relevant range of $z_{cut}$.
 Only the non-BLM  corrections are used in the calculation.
 } \label{q7}
\end{figure}

The next observable we consider is the first moment of the photon
energy spectrum $\langle E _\gamma \rangle$ for $E_\gamma > E_{\rm
cut}$. It is straightforward to derive the correction to the
average photon energy when the second order QCD corrections are
applied. The result reads:
\begin{eqnarray}
&& \delta \langle E_\gamma \rangle =  - {m_b\over 2} \left
(\frac{\alpha_s}{\pi} \right )^2 \Delta_E(z_{\rm cut}),
\nonumber \\
&& \Delta_E(z_{\rm cut}) = C_F \int \limits_{z_{\rm cut}}^{1} {\rm
d} z (1-z) F^{(2)}(z)
\nonumber \\
&& + C_F^2 \int \limits_{z_{\rm cut}}^{1} {\rm d} z (1-z)
F^{(1)}(z) \int \limits_{0}^{z_{\rm cut}} {\rm d} y F^{(1)}(y) .
\label{E1}
\end{eqnarray}
The last term in Eq.(\ref{E1}) comes from the interference of
${\cal O}(\alpha_s)$ corrections to the numerator and denominator
in the expression for $\langle E_\gamma \rangle$.  We disregard
the BLM corrections and compute $\Delta_E$ as a function of
$z_{\rm cut}$. The result is shown
in Fig.\ref{q7}.
For $E_{\rm cut} = 1.8~{\rm GeV}$, we get  $z_{\rm cut}
= 0.78$ and  obtain $\Delta_E = -0.68$. This leads to:
\begin{equation}
\delta \langle E_\gamma \rangle_{\rm non-BLM}
= \left ( 1.8 \pm 0.3 \right ) \times 10^{-2}~{\rm GeV},
\label{eav}
\end{equation}
depending on the value of the strong coupling constant.

The current most accurate measurement of the average energy in $B
\to X_s + \gamma$ comes from Belle collaboration \cite{Belle2004}
and reads $\langle E_\gamma \rangle_{E _\gamma >1.8~\GEV} =
(2.292\pm 0.026 \pm 0.034)~\GEV$. The theoretical estimates for
this observable that are available in the literature are: $\langle
E_\gamma \rangle = 2.27^{+0.05}_{-0.07}~{\rm GeV}$
\cite{Neubert2004}, where only the uncertainty associated with
uncalculated higher order corrections is displayed,  and
$\langle E _\gamma \rangle  = 2.312~\GEV$ \cite{BBU2004} (no explicit
estimate of  the uncertainty has been given in that reference).
Our result Eq.(\ref{eav})
shows that the central values in these predictions should be
shifted approximately by one percent.  In addition,
as the result of our calculation, the error
bars for  $\langle E_\gamma \rangle$
associated with higher order QCD corrections  given in \cite{Neubert2004}
should be reduced  to at most a percent.

We note that although a percent shift in the central value
for $\langle E_\gamma \rangle$ is small, it can not be
considered negligible given the
current precision achieved in $B$-physics. In particular, the
average photon energy plays a central role in constraining the
mass of the $b$-quark from $B$-decays.
The correction that we have computed in Eq.(\ref{eav})
results in {\it approximately} $-30~{\rm MeV}$ shift in the
short-distance low-scale mass. The value of this shift can not be
neglected because it roughly equals the uncertainty in the value
of the $b$-quark mass that was extracted from recent fits to
semi-leptonic and radiative $B$-decays \cite{bbprl}.

\section{Conclusions}

The knowledge of the photon energy spectrum in $B\to X_s+\gamma$
is required to relate the experimental measurements of the
branching fraction ${\rm Br}[B\to X_s+\gamma]$ with a lower
cut on the photon energy and the fully inclusive theoretical
calculations. This observable is of particular interest
because it is sensitive to physics beyond the Standard Model and
represents the cleanest observable for theoretical computations.
In addition, the photon energy spectrum is used to test our
understanding of strong interactions and to extract fundamental
parameters of heavy flavor physics.

In this paper the analytic computation of the
${\cal O}(\alpha_s^2)$ corrections to the photon energy spectrum
is presented.
We have restricted ourselves to the contribution of the operator
$\hat{O}_7$. Partial analytical results on the ${\cal O}(\alpha_s^2)$
corrections to the photon energy spectrum that have been published
previously are confirmed. On the other hand, we have shown that
the  terms that are singular in $z \to 1$ limit do not furnish
a good numerical approximation to ${\cal O}(\alpha_s^2)$ corrections for
$z < 0.95$.

To evaluate the numerical impact of the ${\cal O}(\alpha_s^2)$
corrections, we separate the BLM and the non-BLM pieces. While the
BLM corrections have been computed and analyzed previously, the
non-BLM component at ${\cal O}(\alpha_s^2)$ represents the new
result of this paper. The impact of this new result on the
fraction of events with photons having energies larger than
$E_{\rm cut}$ and on the average energy of the photons in $B \to
X_s + \gamma$ is in the $1\%$ range. This new contribution is
comparable in size to both the accuracy of the experimental
measurements and the uncertainty assigned to these observables in
previous evaluations.

{\bf Acknowledgments.} We are grateful to  N.~Uraltsev for useful
communications and to  T.~Browder for experimental insights.
This research is supported by
the DOE under contract DE-FG03-94ER-40833, the  Outstanding Junior
Investigator Award DE-FG03-94ER-40833 and by the start up funds of
the University of Hawaii.


\end{document}